\documentclass[aps,pra,reprint,groupedaddress]{revtex4-1}

\usepackage{graphicx}
\usepackage{amsmath}
\usepackage{color}

\begin{document}


\title{Loss-based Experimental Test of Bohr's Complementarity Principle with Single Neutral Atom}


\author{Zhihui Wang}
\author{Yali Tian}
\author{Chen Yang}
\author{Pengfei Zhang}
\author{Gang Li}
\email[]{gangli@sxu.edu.cn}
\author{Tiancai Zhang}
\email[]{tczhang@sxu.edu.cn}
\affiliation{State Key Laboratory of Quantum Optics and Quantum Optics Devices, Institute of Opto-Electronics, Shanxi University, Taiyuan 030006, China}
\affiliation{Collaborative Innovation Center of Extreme Optics, Shanxi University, Taiyuan 030006, China}

\date{\today}

\begin{abstract}
An experimental test of quantum complementarity principle based on single neutral atom trapped in a blue detuned bottle trap was here performed. A Ramsey interferometer was used to assess the wavelike behavior or particle-like behavior with second $\pi/2$-rotation on or off. The wavelike behavior or particle-like behavior is characterized by the visibility $V$ of the interference or the predictability $P$ of which-path information, respectively. The measured results fulfill the complementarity relation $P^2+V^2\leq1$. Imbalance losses were deliberately introduced to the stem and find the complementarity relation is then formally ``violated.'' All the experimental results can be completely explained theoretically by quantum mechanics without considering the interference between wave and particle behaviors. This observation complements existing information concerning the BCP based on wave-particle duality of massive quantum.
\end{abstract}


\maketitle

Bohr's complementarity principle (BCP) is one of the cornerstones of quantum mechanics, and the counterintuitive behavior of wave-particle duality lies at its heart \cite{Bohr28}. BCP says that the properties of waves and particles for a quantum system cannot be simultaneously observed. Various tests of BCP with single photons have been performed \cite{Jelezko03,Zeilinger05,Aichele05,Jacques05,Jacques07,Jacques08,Jacques008,Tang12,Tang13}. However, the low detection efficiency associated with fast-moving, massless photons makes the results less persuasive and quite untenable. Here we use a well-controlled, massive, single trapped Cesium atom in a Ramsey interferometer to test BCP of wave-particle duality. A single atom is detected with much greater efficiency and our results confirm the complementarity relation. We also deliberately introduce imbalance losses into our system and find the complementarity relation is formally ``violated''. The whole experiment is closer to the classical notions, and the result is more ideal than ever, which makes BCP seem even more firm. Our observation provides an important complementation to understand the BCP of wave-particle duality. The system paves a way to observe selectively the wave-particle properties on a single quantum level for massive particles.

\begin{figure}
\includegraphics[width=7.5cm]{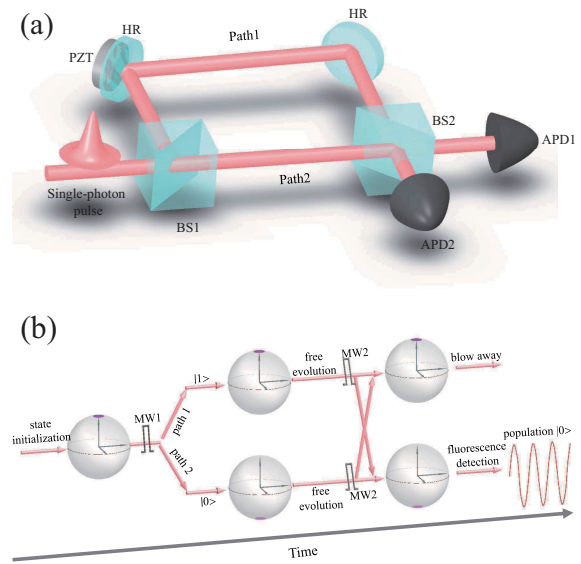}
\caption{\label{fig1} (color online) (a) Schematic of BCP test with single photon in the conventional MZI. A single photon state was transformed into a superposition of two paths, 1 and 2, in space by the first beam splitter BS1. The phase shift $\theta$ of two arms was tuned by scanning the voltage of the piezoelectric transducer (PZT). The presence or absence of the second beam splitter BS2 helped us observe the wave- or particle-like behaviors of single photons. Two detectors, APD1 and APD2, were used to count photons from two output ports. (b) Schematic of the Ramsey interferometer system. A single atom state was split by the first microwave pulse MW1 into superposition states of $\left| 0 \right\rangle$ and $\left| 1 \right\rangle$, which were similar to the two paths and the state evolution in the time domain. After a certain time, the second microwave pulse MW2 was present (or absent) and the final state of the atom could be detected by regular probe-fluorescence measurement.}
\end{figure}

Wheeler's gedanken delayed-choice experiment \cite{Wheeler84,Wheeler78} and the corresponding realizations  \cite{Jacques07,Jacques08,Tang12,Manning15} reveal the nature of the fundamental particles of photons or atoms; they simultaneously possess behaviors of wave-particle duality until the detection arrangement forces them to behave as either one or the superposition of both. In a two-path interferometer, e.g., a Mach-Zehnder interferometer [MZI, see Fig. \ref{fig1}(a)], by moving the second beam splitter (BS) BS2 in or out, we can examine the two exclusive properties of waves and particles, respectively. With the BS2 in the MZI, there is interference between the two paths. By varying the phase difference between these two paths, we can observe an interference fringe, and thus we can observe the pure wave property. When the second BS is moved out, the MZI is open and the two detectors detect the particle from two separate paths. Which-path information is known deterministically, and the photon shows the property of a pure particle. Between these two all-or-nothing cases there also exists an intermediate situation, where by setting the reflectivity of the BS to a value between 0 and 0.5, the two exclusive properties can be partially shown simultaneously. This intermediate case was first theoretically studied by Wooster and Zurek \cite{Wootters79} in 1978, and it was used to test the BCP. A qualitative formulation of BCP based on which-path information and interference visibility was discussed \cite{Wootters79}. Later, in a neutron interferometry experiment, the partial which-path information and limited visibility were observed at the same time \cite{Summhammer83}. In 1995 and 1996, Jaeger et al. and Englert independently obtained an inequality to quantitatively describe the BCP of wave-particle duality, which is \cite{Jaeger95,Englert96}

\begin{equation}
\label{eq:1}
P^2+V^2\leq1.
\end{equation}
Here, $V$ is the visibility of the wave interference and $P$ is the predictability of which-path information.

The experimental examination of inequality (\ref{eq:1}) is firstly done based on a large number of photons from a faint laser \cite{Hellmuth87}, and the result obeys the inequality (\ref{eq:1}) very well. However, the experimental results could also be explained by classical electrodynamics. In 1998, S. D\"urr et al. used an atom interferometer controlled by atomic internal state to test the BCP in a quantum regime \cite{Durr98}. Although plenty of atoms were used in the experiment, the result can only be explained by quantum mechanics, and the results fulfill inequality (\ref{eq:1}). Several experimental tests of BCP on a single-photon level have been executed since 2007 \cite{Jacques07,Jacques08,Jacques008,Tang12,Tang13}. Results are consistent with inequality (\ref{eq:1}), except for one experiment \cite{Tang13}, in which it was observed that $P^2+V^2>1$. This abnormality was attributed to interference between wave and particle behaviors. So far, all the experimental tests of BCP with single photons, as well as experiments with atoms [19], have suffered from limited detection efficiency, which implies that only some of the photons were registered and used to eventually evaluate the results. This makes the results less persuasive and less tenable. In our experiment, we performed a BCP examination with a detection efficiency of 0.75, with a single trapped neutral atom in a Ramsey interferometer [see Fig. \ref{fig1}(b)]. Our experimental results fulfill inequality (\ref{eq:1}) quite well. When imbalance losses are deliberately introduced into the Ramsey interferometer, we find that the complementarity relation is formally ``violated''. In a previous experiment we also observed imbalance-induced ``violation'' with a faint laser in MZI \cite{Yang13}. It can be theoretically explained by quantum mechanics without considering the interference between the wave and particle behaviors. The ``violation'' is simply due to the imbalance losses in the two arms of the interferometer. Our observation provides an important complementation to understand the BCP of wave-particle duality.

In our experiment, we used a well-designed Ramsey interferometer for single trapped Cesium atoms, where the two atomic wave packages referring to $\left| 0 \right\rangle  = \left| {F = 3,{m_F} = 0} \right\rangle$ and $\left| 1 \right\rangle  = \left| {F = 4,{m_F} = 0} \right\rangle$ states represent two material wave paths. The atom was initially prepared in state $\left| 1 \right\rangle$ with high efficiency ($>$0.99), and two $\pi/2$ pulses driven by microwave fields with frequency resonating to the atomic transition $\left| 0 \right\rangle \leftrightarrow \left| 1 \right\rangle$ were applied sequentially to separate the wave package into the two paths and combine them again. The two $\pi/2$ pulses are functionally equivalent to the two beam splitters in MZI. By changing the length of the microwave pulse away from $\pi/2$, we could partially separate and combine the two paths, similar to the ratio change of the beam splitter in MZI, and this enabled us to test the BCP in the intermediate regime.

\begin{figure}
\includegraphics[width=7.5cm]{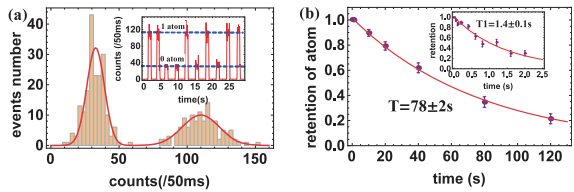}
\caption{\label{fig2} (color online) (a) Typical signal from single atoms: fluorescence recorded by the single photon counting module (inset) and the corresponding histogram. (b) The average lifetime of a single atom in the blue trap was about 78 s, and the internal state lifetime for the trapped atoms in $\left| 0 \right\rangle$ and $\left| 1 \right\rangle$ was about 1.4 s (inset).}
\end{figure}

A single cesium atom was isolated from a conventional magneto-optical trap (MOT) using a blue detuned bottle trap \cite{Wang15} by light-assisted collisions \cite{Schlosser01,Wang14}. The trap had a size of about 11 $\mu m$ in the axial direction and 2 $\mu m$ in the radial directions. The trap was sufficiently small in size that the system was operating in the collisional blockade regime \cite{Schlosser02,Anderson10}, which ensured that no more than one atom could be loaded into the trap. The typical fluorescence signals of single atoms are shown in Fig. \ref{fig2}(a). The probability that two atoms would be loaded in the trap simultaneously is zero. That is why it is a single atom system. The trapped atom was cooled to a temperature of about 10 $\mu K$ by polarization gradient cooling, and the corresponding de Broglie material wave length was about 69 nm. So our experiment of the BCP test of wave-particle duality was strictly performed on a single atom, which showed classical behavior. The exponential lifetime for the trapped atom was 78 s, and the internal state lifetime for the trapped atoms for $\left| 0 \right\rangle$  and $\left| 1 \right\rangle$  was over 1 second [see Fig. \ref{fig2}(b)]. The overall execution time for the Ramsey interferometer was less than 500 $\mu s$, thus the atom loss rate and state damping rate over the interferometer executing time were extremely low (only 0 and 0.0004, respectively).

The state detection of the atom at the output of the Ramsey interferometer was accomplished by the usual fluorescence detection technique \cite{Kuhr03,Yavuz06}. The resonant probe light pulse to $\left| {6{S_{1/2}}F = 4} \right\rangle  \leftrightarrow \left| {6{P_{1/2}}F = 5} \right\rangle$ was applied, and the scattered photons were collected and detected by single photon detectors. For those atoms in path $\left| 0 \right\rangle$, the detector would get the scattering photons, while there were no scattering photons if the atoms were in path $\left| 1 \right\rangle$. Atoms in these two paths could be distinguished with discrimination of over 0.99 [see Fig. \ref{fig2}(a)]. However, due to the heating, about 0.25 of atoms were lost from the blue trap before enough photons were collected in the detection process. So the overall atom detection efficiency was about 0.75, including the transmission efficiency of the Ramsey interferometer, which was still the best in such a BCP test experiment with single particles.

\begin{figure}
\includegraphics[width=7.5cm]{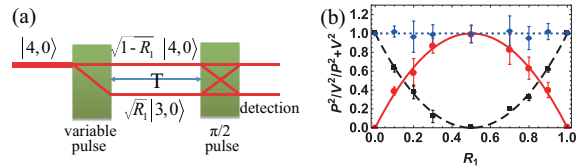}
\caption{\label{fig3} (color online) Test of BCP with single cesium atom. (a) Schematic of BCP relation test with single atom Ramsey interferometer. Single atom initially in state $\left| 1 \right\rangle$ was subjected to two rotation operations created by microwave pulse analog of the beam splitters in the usual MZI. In presence of the second microwave pulse, the probability of measuring the qubit in a certain state depends on the relative phase $\theta$, thus exhibiting wavelike behavior, while in the absence of the pulse, the probability is independent of $\theta$ and particle-like behavior appears. (b) Wavelike information ${V^2}$, which-path information ${P^2}$ and ${P^2} + {V^2}$ as functions of $R_1$, which is determined by the length of the first microwave pulse. The red dots, black squares, and blue diamonds stand for the measured values of ${V^2}$, ${P^2}$ and ${P^2} + {V^2}$ without losses, and the red solid, black dashed and blue dotted lines are theoretical fittings according to equations (\ref{eq:2}-\ref{eq:4}).}
\end{figure}

We first varied the length of the first microwave pulse from 0 to $\pi$, corresponding to the reflective coefficient $R_1$ of the first beam splitter BS1 from 0 to 1, and fixed the length of the second microwave pulse at $\pi$/2, corresponding to a reflectivity coefficient 0.5 of the second beam splitter BS2 [see Fig. \ref{fig3}(a)]. For a given value $R_1$, the wavelike information of the single atoms was obtained by measuring the visibility $V$ of the interference observed by scanning the time interval $T$ between the two microwave pulses, which is \cite{Jia14,Jia014}
\begin{equation}
\label{eq:2}
V = 2\sqrt {{R_1}\left( {1 - {R_1}} \right)}.
\end{equation}
On the other hand, the predictability $P$ is required to qualitatively characterize which-path information and then test the BCP inequality. The predictability is defined as
\begin{equation}
\label{eq:3}
P = \left| {1 - 2{R_1}} \right|.
\end{equation}
We thus get the relation
\begin{equation}
\label{eq:4}
{P^2} + {V^2} = 1.
\end{equation}
which is independent of $R_1$. The measured results are depicted in Fig. \ref{fig3}(b), which shows ${P^2}$, ${V^2}$ and ${P^2} + {V^2}$ as functions of $R_1$. The lines are the theoretical fittings \cite{Jia014} and clearly the results fulfill the inequality ${P^2} + {V^2} \le 1$.

In order to investigate the influence of the imbalance of a Ramsey interferometer due to the losses on the BCP relation, we deliberately introduced a controllable loss only on path 2 by coupling certain members of the population in $\left| 1 \right\rangle$ out of the interferometer. The controllable loss was introduced by applying a mixed $\pi$/2 Raman pulse composed of a $\pi$-polarized light beam and a ${\sigma ^{\rm{ + }}}$-polarized light beam at 894nm. The single-photon detuning of a Raman pulse was -35 GHz away from the D1 transition. This Raman beam was two-photon resonant to the transition frequency from $|\left. {6{S_{1/2}}F = 3,{m_F} = 0} \right\rangle $ to $|\left. {6{P_{1/2}}F = 4,{m_F} = -1} \right\rangle $. When the losses occur inside the interferometer, equations (\ref{eq:2}-\ref{eq:4}) become the following:

\begin{equation}
\label{eq:5}
V = \frac{{2\sqrt {{R_1}\left( {1 - {R_1}} \right)\left( {1 - {L_1}} \right)\left( {1 - {L_2}} \right)} }}{{\left( {1 - {R_1}} \right)\left( {1 - {L_1}} \right) + {R_1}\left( {1 - {L_2}} \right)}}.
\end{equation}
\begin{equation}
\label{eq:6}
P = \frac{{\left| {\left( {1 - {R_1}} \right)\left( {1 - {L_1}} \right) - {R_1}\left( {1 - {L_2}} \right)} \right|}}{{\left( {1 - {R_1}} \right)\left( {1 - {L_1}} \right) + {R_1}\left( {1 - {L_2}} \right)}}.
\end{equation}
\begin{equation}
\begin{aligned}
\label{eq:7}
{P^2} + {V^2} & = \frac{{4R\left( {1 - {R_1}} \right)\left( {1 - {L_1}} \right)\left( {1 - {L_2}} \right)}}{{{{\left[ {\left( {1 - {R_1}} \right)\left( {1 - {L_1}} \right) + {R_1}\left( {1 - {L_2}} \right)} \right]}^2}}} \\
&+ \frac{{{{\left[ {\left| {\left( {1 - {R_1}} \right)\left( {1 - {L_1}} \right) - {R_1}\left( {1 - {L_2}} \right)} \right|} \right]}^2}}}{{{{\left[ {\left( {1 - {R_1}} \right)\left( {1 - {L_1}} \right) + {R_1}\left( {1 - {L_2}} \right)} \right]}^2}}}.
\end{aligned}
\end{equation}

When the losses occur outside the interferometer, visibility $V$ and predictability $P$ can also be found:

\begin{equation}
\label{eq:8}
V = 2\sqrt {{R_1}\left( {1 - {R_1}} \right)} .
\end{equation}
\begin{equation}
\label{eq:9}
P = \frac{{\left| {\left( {1 - {R_1}} \right)\left( {1 - {L_1}} \right) - {R_1}\left( {1 - {L_2}} \right)} \right|}}{{\left( {1 - {R_1}} \right)\left( {1 - {L_1}} \right) + {R_1}\left( {1 - {L_2}} \right)}}.
\end{equation}
\begin{equation}
\label{eq:10}
\begin{aligned}
{P^2} + {V^2} & = \frac{{{{\left[ {\left| {\left( {1 - {R_1}} \right)\left( {1 - {L_1}} \right) - {R_1}\left( {1 - {L_2}} \right)} \right|} \right]}^2}}}{{{{\left[ {\left( {1 - {R_1}} \right)\left( {1 - {L_1}} \right) + {R_1}\left( {1 - {L_2}} \right)} \right]}^2}}} \\
&+ 4{R_1}\left( {1 - {R_1}} \right).
\end{aligned}
\end{equation}

\begin{figure}
\includegraphics[width=7.5cm]{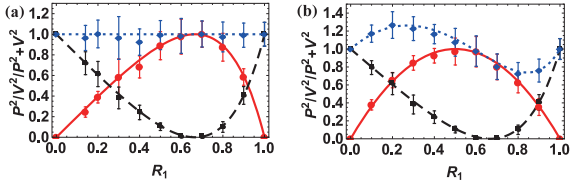}
\caption{\label{fig4} (color online) The results of BCP relation test when losses were introduced. (a) The results for losses occurring inside the interferometer. (b) The results for losses occurring outside the interferometer. The red dots, black squares, and blue diamonds stand for the measured values of ${V^2}$, ${P^2}$ and ${P^2} + {V^2}$ with losses ( $L_1=0$, $L_2=0.5$) in path 1 and path 2, respectively. The red solid, black dashed and blue dotted lines are theoretical fittings according to equations (\ref{eq:5}-\ref{eq:10}).}
\end{figure}

Figure \ref{fig4} shows the experimental results, for which the losses on paths 1 and 2 are $L_1=0$ and $L_2=0.5$, respectively. It indicates that both ${V^2}$ and ${P^2}$ are no longer symmetrical, but the BCP relation ${P^2} + {V^2} = 1$ still holds when the losses occur inside the interferometer. Losses occurring inside the interferometer influence both the visibility and predictability, but have no effect on the BCP relation. Figure \ref{fig4}(b) gives the results for the case when losses occur after the second $\pi$/2 pulse (outside the interferometer) and all other settings remain unchanged. In this situation ${P^2}$ is not symmetrical and ${V^2}$ does not change. It is obvious that the measured BCP inequality ${P^2} + {V^2} \le 1$ no longer holds. Here, losses have no effect on the visibility but have influences on the predictability. The BCP relation looks ``violated.'' This is because some of the atoms are lost and the measured predictability is not the originally defined predictability \cite{Englert96}.

\begin{figure}
\includegraphics[width=7.5cm]{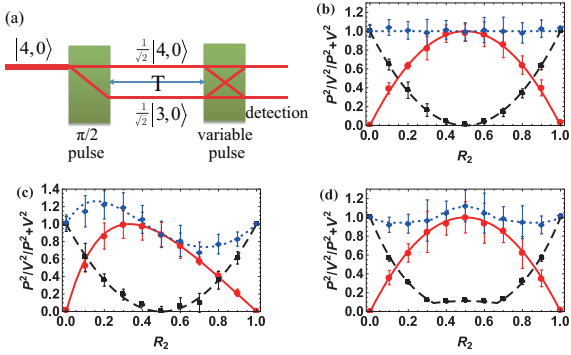}
\caption{\label{fig5} (color online) Test of BCP relation of the second configuration. (a) Schematic of BCP relation test with Ramsey interferometer where the second pulse was varied. (b) Experiment results without losses. Experiment results with losses occurring inside (c) and outside (d) the Ramsey interferometer. The red dots, black squares, and blue diamonds stand for the measured values of ${V^2}$, ${P^2}$ and ${P^2} + {V^2}$ and the red solid, black dashed and blue dotted lines are theoretical fittings according to equations (\ref{eq:11}-\ref{eq:16}).}
\end{figure}

In order to comprehensively investigate how the losses affected the BCP relation, we further designed the second configuration, as shown in Fig. \ref{fig5}(a). We kept the first microwave pulse as a fixed $\pi$/2, corresponding to the reflective coefficient 0.5 for the first BS1, and varied the length of the second microwave pulse from 0 to $\pi$, corresponding to the reflective coefficient $R_2$ varying from 0 to 1. In the lossless case, the visibility $V$ and predictability $P$ are determined by equations (\ref{eq:2}) and (\ref{eq:3}). The experimental results are shown in Fig. \ref{fig5}(b), which is the same as in Fig. \ref{fig3}(b). For losses occurring inside the interferometer in the second configuration, $V$ and $P$ become

\begin{equation}
\label{eq:11}
V = \frac{{2\sqrt {{R_2}\left( {1 - {R_2}} \right)\left( {1 - {L_1}} \right)\left( {1 - {L_2}} \right)} }}{{\left( {1 - {R_2}} \right)\left( {1 - {L_1}} \right) + {R_2}\left( {1 - {L_2}} \right)}}.
\end{equation}
\begin{equation}
\label{eq:12}
P = \left| {1 - 2{R_2}} \right|.
\end{equation}
\begin{equation}
\begin{aligned}
\label{eq:13}
{P^2} + {V^2} &= \frac{{4{R_2}\left( {1 - {R_2}} \right)\left( {1 - {L_1}} \right)\left( {1 - {L_2}} \right)}}{{{{\left[ {\left( {1 - {R_2}} \right)\left( {1 - {L_1}} \right) + {R_2}\left( {1 - {L_2}} \right)} \right]}^2}}} \\ &+ {\left( {1 - 2{R_2}} \right)^2}.
\end{aligned}
\end{equation}

For losses occurring outside the interferometer, the results are as follows:

\begin{equation}
\label{eq:14}
V = 2\sqrt {{R_2}\left( {1 - {R_2}} \right)}.
\end{equation}
\begin{equation}
\label{eq:15}
\begin{aligned}
P & = \frac{{\left| {\left( {1 - {R_2}} \right)\left( {1 - {L_1}} \right) - {R_2}\left( {1 - {L_2}} \right)} \right|}}{{2\left( {1 - {R_2}} \right)\left( {1 - {L_1}} \right) + {R_2}\left( {1 - {L_2}} \right)}} \\
&+ \frac{{\left| {\left( {1 - {R_2}} \right)\left( {1 - {L_2}} \right) - {R_2}\left( {1 - {L_1}} \right)} \right|}}{{2\left( {1 - {R_2}} \right)\left( {1 - {L_2}} \right) + {R_2}\left( {1 - {L_1}} \right)}}.
\end{aligned}
\end{equation}
\begin{equation}
\label{eq:16}
{P^2} + {V^2} = {P^2} + 4{R_2}\left( {1 - {R_2}} \right).
\end{equation}

Figures \ref{fig5}(c) and (d) show the results corresponding to cases where the losses were introduced inside and outside the Ramsey interferometer, respectively. The red solid, black dashed, and blue dotted lines are theoretical fittings by equations (\ref{eq:11})-(\ref{eq:16}), and it is clear that the measured results are in good agreement with the theory. The visibility ${V^2}$, probability ${P^2}$ and ${P^2} + {V^2}$ are again shown in Fig. \ref{fig5}(c) and (d). The behavior in Fig. \ref{fig5}(c) is very similar to the result in Fig. \ref{fig4}(b), with losses occurring inside the Ramsey interferometer, and we again find that the BCP relation is ``violated'' due to breaking of the balance between the two arms of the Ramsey interferometer. For the case with losses outside the Ramsey interferometer, see Fig. \ref{fig5}(d), whose results are symmetrical. We find that ${V^2}$ is independent of the losses, while ${P^2}$ and ${P^2} + {V^2}$ have two turning points. The BCP test results show a striking straw-hat shape, and the ``violation'' of the inequality is observed when $R_2$ is around 0.5. All these weird-looking properties are explained exactly by our theoretical calculations \cite{Jia14,Jia014} and are consistent with our previous results with photons \cite{Yang13}.

\begin{figure}
\includegraphics[width=7.5cm]{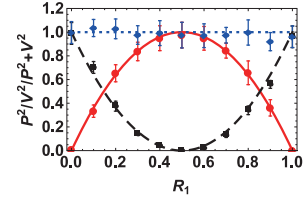}
\caption{\label{fig6} (color online) Elimination of the influence of the imbalance loss for the configuration with loss outside the Ramsey interferometer [Figure (\ref{fig4})(b)]. The ``violation'' is eliminated by switching the input state and then averaging the results. The red dots, black squares, and blue diamonds stand for the measured values of ${V^2}$, ${P^2}$ and ${P^2} + {V^2}$ respectively, and the red solid, black dashed and blue dotted lines are theoretical fittings according to equations (\ref{eq:2}-\ref{eq:4}).}
\end{figure}

We would like to point that the ``violation'' is simply due to imbalance between two paths with losses introduced either inside or outside the Ramsey interferometer. In principle this ``violation'' can be eliminated by switching the input state from $\left| 1 \right\rangle$ to $\left| 0 \right\rangle$ (switching the two paths) and then averaging the results. The corrected experimental results are given in Fig. \ref{fig6}, from which we can see that the BCP relation ${P^2} + {V^2} \le 1$ is attained.

In conclusion, we tested the BCP relation by wave-particle duality measurement with an elaborately designed Ramsey interferometer based on well-controlled single neutral Cesium atoms trapped in a blue bottle trap. We proved that the BCP inequality ${P^2} + {V^2} \le 1$ is validated without losses in a Ramsey interferometer. Two experimental configurations, in which the extra losses were introduced either inside or outside the Ramsey interferometer, are demonstrated. All the observed BCP features characterized by the sum of wave- and particle-like information ${P^2} + {V^2}$, along with the reflectivity of the beam splitter, are well explained. The observed ``violation'' of the BCP inequality ${P^2} + {V^2}>1$ is simply due to the imbalance between the two arms of the interferometer. This seeming ``violation'' can be completely eliminated by switching the two paths and then averaging the results. The corrected results prove that BCP is still valid. Our experimental test of BCP is more ideal than ever because: 1) it is based on single, deterministic, massive atoms; and 2) the detection efficiency of single atoms is much higher than that of experiments with photons, which make the result more persuasive. The experiment has paved the way to selectively observing the wave and particle properties on a single quantum level for massive particles.

\begin{acknowledgments}
We would like to thank S. Y. Zhu and A. A. Jia for helpful discussions. This research is supported by the National Basic Research Program of China (Grant No. 2012CB921601), NSFC (Grant No. 11634008, No. 11674203, No. 61227902, No. 91336107, and No. 61275210).
\end{acknowledgments}

\bibliography{BCPtest}

\end{document}